# Rain energy harvesting using atomically thin Gadolinium Telluride decorated 3D Printed nanogenerator

Partha Kumbhakar[1§], Arko Parui[2§], Rushikesh S. Ambekar[1], Madhubanti Mukherjee[2], Saif Siddique[1], Nicola M. Pugno[3,4], Abhisek K. Singh[2*], Chandra S. Tiwary[1]*

[1]Metallurgical and Materials Engineering, Indian Institute of Technology Kharagpur, Kharagpur-721302, India
[2]Materials Research Centre, Indian Institute of Science, Bangalore, Karnataka 560012, India
[3]Laboratory of Bio-inspired, Bionic, Nano, Meta Materials and Mechanics, Department of Civil, Environmental and Mechanical Engineering, University of Trento, Italy
[4]School of Engineering and Materials Science, Queen Mary University of London, Mile End Road, London E1 4NS, UK
*
E-mail: abhishek@iisc.ac.in, chandra.tiwary@metal.iitkgp.ac.in

**Abstract:** The 3D printing technology offers an innovative approach for developing energy storage devices to create facile and low-cost customized electrodes for modern electronics. Generating electric potential by moving a droplet of ionic solution over two-dimensional (2D) materials is a novel method for rain energy harvesting. This work demonstrated a liquid-solid contact electrification-based 3D printed nanogenerator where raindrop passes through the positively charged ultrathin Gadolinium Telluride ($Gd_2Te_3$) sheets. Experimental results showed that voltage as high as ~0.6 V could be generated by moving a droplet of ionic solution on the decorated 3D printed nanogenerator. The output efficiency of the nanogenerator is increased ~400% by enhancing the surface area of copious 3D printed porous structures. Density Functional Theory (DFT) calculations are done, revealing that the high electrical conductivity of (112) surface of $Gd_2Te_3$ is due to the p-type charge carriers. Additionally, we illustrate the enhancement of the output performance (~0.8V) by using a graphite rod and arbitrarily manipulating the surface charge. Therefore, this work can open up a new avenue to advance scientific research of Blue energy harvesting and tackle the energy crisis.

**Keywords**: Telluride, 2D materials, 3D printing structure, Energy harvesting, DFT.

# 1. Introduction

The limited availability of fossil fuels and their negative environmental impacts has driven the research for sustainable technologies to meet current energy challenges [1]. Hydropower-based energy seems to be the only contender to replace traditional energy generation systems among the existing technologies. With the abundance of water on the earth's crust, generating electricity from flowing water is still an enticing opportunity. If we consider the example of rainfall, an enormous amount of energy can be harvested from it if the energy of each drop is successfully converted to electrical output and stored [2]. Each raindrop carries energy in two forms: the kinetic energy when they fall and the electrostatic energy due by friction with other water droplets or air molecules [3] [4]. New and effective technology is required to convert and store this energy efficiently. Researchers are made smart materials capable of converting the kinetic energy of liquid into electricity have also been developed [5] [6] [7]. There are two techniques for inducing local drop motion: (a) direct contact, (b) contactless technique. Contactless methods have several advantages, such as loss of liquid mass, surface contamination, and manipulation of the desired precision of the liquid drop. They involve electrowetting, dielectrophoresis, and charge printing [8]. Innovative techniques such as using magnetic fields variations and light-induced capillary forces to tubular microactuators to propel liquid have also been suggested [9]. Recently, 3D printing-based flexible, efficient, and economical energy storage devices are gained huge popularity [10] [11] due to their several advantages such as flexibility, geometry controllability, lightweight, customizable, high mechanical property, single printing system for complex or different materials, etc.[12] [13] [14]. Since the amount of energy generation directly depends on contact surface area therefore 3D printing can be a crucial techinque as it can easily be utilized for the fabrication of complex porous structures that can provide higher surface area [15] [16]. Building 3D printed structures of atomically thin (2D-two dimensional) materials of high surface area and unique physical and chemical properties opens up a new advancement for making flexible electrode materials.

2D Tellurides have attracted much attention in the past few decades due to their topological semimetal nature magnetic properties and superconducting behavior [17] [18]. When rare earth metals are alloyed with tellurides, they add to the already rich physical and chemical properties of tellurides. [19] [20] The large surface area offered by 2D material makes it a promising material for such drop motion studies [7]. Additionally, only the top layer, in contact with the liquid, generates voltage; the other layers act as shunt resistors dividing the output voltage when the system is considered a nanogenerator circuit. Therefore, the lesser the number of layers more excellent the output voltage [21]. The Gd-Te system has consisted of two sites for Gd and three for Te; it does not have any cation vacancies or sites for other atoms [22] [23]. Because of its similarity with tetradymite structures ($A_2X_3$; A = Bi, Sb, and X = Se, Te), Gd-chalcogenides show the best thermoelectric properties among all the rare-earth chalcogenides, with $Gd_2Te_3$, in particular, having a power factor of 0.6 X $10^{-6}$ W $K^{-1}$ $cm^{-1}$ [23]. The material was found to exhibit semi-metallic behavior and an antiferromagnetic magnetic ground state. The present work reports the yet unexplored physical properties of 2D $Gd_2Te_3$ as energy harvesting materials. The ultrathin $Gd_2Te_3$ decorated 3D-printed structures act as a flexible device for energy harvesting by manipulating drops of ionic liquid on its active surface, as shown in **Figure 1**. The atomically thin material decorated 3D printed structure enhances the overall performance by contact electrification and impulse by raindrops. Different types of 3D printed porous structures are employed to study the effect of porosity and effective surface area in energy generation. A detailed Density Functional Theory (DFT) calculations were also done to insight the origin of generating voltage by $Gd_2Te_3$ coated 3D printed structure and their nature of charge carrier.

## 2. Results and Discussions

**Figure 2A** shows the XRD pattern of bulk $Gd_2Te_3$. The pattern matches the *Pnma* phase with a $Sb_2S_3$ (stibnite) structure. Inset shows the digital photograph of the bulk $Gd_2Te_3$ crystal with a size of a few centimeters. These crystals were extremely brittle and showed a planar fracture surface as observed by scanning electron microscopy (SEM) (**Figure S1**). Composition analysis was performed using Energy Dispersive X-ray Spectroscopy (EDX) to ensure that proper composition had been achieved (**Figure S2**). The results for the selected area show that the sample has 56.56% of Te and 43.44% Gd by mass. $Gd_2Te_3$ is made up of planar sheets stacked on top of one another, making it a layered crystal. Ultrathin layers of $Gd_2Te_3$ samples were synthesized from these bulk crystalline solids, which were powdered initially and then exfoliated with an organic solvent, isopropyl alcohol (IPA) (more details, see Experimental section). This enables us to obtain a dispersion of 2D sheets by adding $Gd_2Te_3$ powder in IPA and agitating the mixture using a probe-sonicator for 30 minutes at room temperature. The vibrational energy provides energy to overcome the van der Waals forces binding the layers and separating them. The XRD was carried out on 2D $Gd_2Te_3$ with Cu-K$\alpha$ radiation ($\lambda = 1.54$ Å) in the 2$\theta$ range of 10-90º (**Figure 2B**). The pattern matches with reported literature and the lattice parameters of $a = 11.96$ Å, $b = 4.29$ Å, $c = 11.75$ Å, $\alpha = \beta = \gamma = 90°$ [23]. The stibnite structure gives the telluride parallel ribbons of $(Gd_4Te_6)_n$ stacked together by weak van der Waals forces, perpendicular to the *b*-axis [24]. The ribbons are made up of Gd-Te-Gd chains, with Te atoms bonded together at 120°, forming of honeycomb-like structure similar to that of graphene. The digital photograph of exfoliated dispersion of $Gd_2Te_3$ at room temperature confirms the formation of ultrathin sheets (inset of **Figure 2B**). Additionally, it can be observed that the XRD peaks of 2D $Gd_2Te_3$ are broader than bulk due to nanostructuring. Also, the overlapping of peaks makes it challenging to distinguish them, and it seems that the number of observed peaks has decreased [25]. The normalized Raman spectra ($\lambda_{ex}$ =532 nm) and microscopy images of 2D $Gd_2Te_3$ of varying thickness are shown in **Figure 2C** and its inset.

The distinct Raman peaks were observed at 100 and 125 cm$^{-1}$. All the spectra had the same peak at 100 cm$^{-1}$. The significant difference observed was the emergence of a peak at 125 cm$^{-1}$ as the thickness of the nanosheets was varied [26]. The origin of these modes is still unclear, and further investigation is needed to explain this confidently. Atomic force microscopy (AFM) results of the exfoliated Gd$_2$Te$_3$ show clear triangular sheet-like structures with a height profile of ~4 nm, corresponding to ~3-4 monolayers (**Figure 2D**). Transmission electron microscopy (TEM) was employed to gain insights into the morphology of the exfoliated 2D sheets. In **Figure 2E,** we present the bright-field TEM image of the exfoliated sample. The images depict atomically thin triangular sheets agreeing with the triangular units found with AFM results. The EDX result confirms the consistency of the atomic ratio of Gd and Te after exfoliation (**Figure 2F**). The dispersion of exfoliated samples was also analyzed in a particle analyzer, and the average particle size obtained was around 450 – 500 nm (**Figure 2G**). Thus, the exfoliation results in sufficiently large sheets with thickness in the order of a few layers. We have performed XPS measurements to get more information about the oxidation states and chemical composition of the exfoliated sample (**Figure S3** and **Figure 2H** and **I**). The XPS analysis shows that the oxidation state of Gd and Te are well-matched with the previous report with proper atomic ratio. The dispersion of ultrathin sheets in IPA was tested for stability using UV-Vis absorption spectroscopy after seven days. More the concentration of these 2D sheets, the higher is the intensity of the UV-Vis spectrum. From **Figure S4,** it is seen that ultrathin Gd$_2$Te$_3$ exhibits similar nature with higher optical absorption in the ultraviolet region than in the visible region of the spectrum. For most of the spectra, the absorption of 2D Gd$_2$Te$_3$ is sufficiently larger, indicating minimum agglomeration and good dispersion stability. It is observed that the spectrum does not have apparent peaks, instead of a broad shoulder. Lai et al. have demonstrated that this indicates the dispersion having few-layer thick nanosheets [27].

To study the performance of electricity generation from ultrathin Gd$_2$Te$_3$ using liquid drop motion, we have fabricated Gd$_2$Te$_3$ based nanogenerator on a 3D printed porous structure.

For the fabrication of nanogenerator (**Figure 1** and see Experimental section for more details), a thin layer of 2D $Gd_2Te_3$, mixed with a PVDF binder in 10:1 ratio, was coated on a 3D printed porous surface. Following this, the nanogenerator was dried on a hot plate for 2 h at 50°C. Copper metal contacts were placed on the positive and negative electrodes at each end of the material. In **Figures 3A** and **3C**, we present the optical microscopy and SEM images of the surface of the 3D printed structure before and after the decorating of the materials. Triangular sheets are well decorated on the polymer substrate. The elemental mapping and EDX spectra also confirm the uniform distribution of the sheets on a 3D printed object (**Figure S5**). The digital photograph of the flexible 3D printed device is depicted in the inset of **Figure 3A**. We also presented the 3D surface plot of the substrates, as shown in **Figures 3B** and **3D**. The surface analysis indicates that after coating with thin sheets, the surface of PLA polymer becomes rough. This will increase the efficacy in the contact area between droplets and materials.

Solid-liquid contact electrification of the 2D $Gd_2Te_3$ material was tested with different sodium chloride (NaCl) solution concentrations. The output voltage was observed at the end of each electrode connected to a digital oscilloscope. For a standard test, the optimized experimental parameters are volume of NaCl droplet is ~20 µL falling from a distance of 6 cm on the 3D printed nanogenerator, which is inclined at an angle of ~15°. The induced output voltage strongly depends on the concentration of the ions present in the droplets. To measure the dependence of output voltage on the ion concentration of brine solution, pure DI water, 0.5 M NaCl, and 1 M NaCl solutions were allowed to flow on the nanogenerator in a static position. The compared voltages are as shown in **Figure 3E**. Interestingly, there is a linear increase in the voltage response as NaCl concentration in water increases. A voltage response for the movement of the liquid droplet on the triangular 2D $Gd_2Te_3$ was measured (**Figure 3F**). The output voltage shows a sharp response to the movement of the droplet. When the droplet moves at a constant velocity on the device, an output voltage of around ~0.5 V was generated in a different direction, i.e., the positive and negative electrodes of the device. This leads to a charge redistribution in the droplet generating electric signals that pass through the external circuit. The output voltage fluctuation is due to the non-uniform distribution of the triangular 2D $Gd_2Te_3$ with the 3D printed substrate. **Figure S6** shows that the static droplet at rest position forms a contact angle of ~90° on the 3D printed device; as the droplet moves downwards on the surface of the cell, the shape of the droplet changes.

To test the feasibility of as-developed setup as an energy harvesting device, multiple droplets of 1M NaCl solution were moved onto a ~15° titled cell, dropped from a height of 6 cm. Output voltages in the range of ~0.4 to ~0.6 V were recorded (**Figure 3G and Movie S1**). We can thus see that the setup works perfectly for a continuous flow of droplets without losing any efficiency. The inset of **Figure 3G** shows a schematic representation of multiple droplet movement directions on the 3D printed cell. The layers of 2D $Gd_2Te_3$ are distinctively shown to represent the charge interaction in between the layers. To assess the effect of sliding angles on the output voltage, we measured output voltage by inclining the 3D printing cell in different angles. We start from 5° of inclination, and as we reach 15°, we get the maximum voltage. This indicates that the output voltage linearly increases with the increasing velocity of the ionic drop. As the sliding angle increases, the contact angle also increases (**Figure S7**), covering a larger surface area and producing more output voltage. The decorated 3D printed device is inclined at an angle of ~15°, giving the droplet a smaller pace in movement. Because a larger surface is being covered, more charges interact on the material junction and drop, generating a larger voltage. Due to the inclination, the side of the drop moving towards an electrode gains more negative charges than the side facing away. We also studied the effect of droplet height fall in voltage generation (**Figure 3H**). We calculate the potential energy ($U_{max}$) by falling drop on the coated substrate. A linear increase in the output voltage up to ~0.5 V in the range of 2.35 to 11. 76× $10^{-7}$ Jule. The inset of **Figure 3H** shows a pictorial representation of the droplet height impacted on the surface, and the cell was inclined at an optimal inclination angle of 15°. We further advanced the application of a 2D $Gd_2Te_3$ decorated 3D printed nanogenerator for energy harvesting by utilizing the dynamics of the ionic solution by applying a graphite rod. We used pencil graphite as conducting substrate. First, we rubbed the graphite and then applied it in the experiment. An increase in the output voltage of the 2D $Gd_2Te_3$ surface with NaCl droplet was observed when a layer of graphite was bought near the surface (**Figure 3I-J**). As the graphite surface touches the NaCl solution, an output voltage is induced repeatedly (**Figure 3J**). It enhanced the overall performance of the cell and showed an increased output voltage of ~0.8 V for multiple droplet movement on the cell. The possible mechanism of the generation of output voltage on $Gd_2Te_3$ surfaces is discussed schematically in **Figure 3I**. During the experiment, more surface charges are obtained on the droplet due to the contact of the graphite rod and $Gd_2Te_3$ substrate. Graphite with multiple graphene layers is supposedly producing static charges, which amplifies the surface charges. The electrostatic charges from a graphite pencil are responsible for the enhanced charge production in the 2D $Gd_2Te_3$ cell. In **Figure 3K**, we have compared the generation of output voltage per unit area ($\xi$) for other 2D materials. It shows that atomically thin $Gd_2Te_3$ decorated 3D printed device shows enhanced performance as compared to other 2D materials [28][29][30][31][32].

The generation of charge due to the contact electrification can be calculated using the formula [3], $Q = 2eN_AclA$, where $e$ is the primary charge, $N_A$ is Avogadro constant, $c$ is the concentration of ions in deionized water/solution, $l$ is the thickness of the ion diffusion, $A$ is the contact area between droplet and $Gd_2Te_3$. Therefore, the charge generation strongly depends on the effective contact surface area. To study the effect of contact specific surface area between the droplets and 3D printed structure, a series of 3D printed structures with different geometry were printed such as hexagonal, line, triangular, and random porosity, and their surface porosity are 33.09%, 39.62%, 16.53%, and 68.71%, respectively. The top and isometric views of these 3D printed structures are illustrated in **Figure 4A** and **Figure 4B,** respectively. To investigate the impact of geometrical property of 3D printed objects, 2D $Gd_2Te_3$ were uniformly coated on them using a similar coating method. Other parameters such as droplet size, solution concentration, dropping speed, and substrate temperature are fixed to compare the output voltage. In **Figure 4C**, we demonstrate the output voltage generation by NaCl droplet. It shows that output voltage increases and gives the highest output for random porosity structures. The output voltage of random porous structure (~0.32 V) enhanced ~400% compared to the solid structure (~0.08 V) owing to random porosity structure (~143 mm$^2$) having higher surface area than solid structure (~59 mm$^2$). The single water droplet of 5 mm diameter has ~59 mm$^2$ surface area; therefore, we have normalized the surface area of all the printed structures by a single droplet contact area. As shown in **Figure 4D**, the specific contact area is increased with an increase in the complexity of the 3D printed structures. These results can be explained as the surface complexity increases, the porosity also changes, leading to the enhancement of output voltage. However, in the case of a solid substrate (no porosity), the output voltage generation is very low compared to the other structures. Moreover, the high random porosity and complex contact surface area of 3D printing structures allow us to a generation of electricity with increased flexibility by tuning the microarchitecture of the 3D printed structures. To demonstrate the real-life application of doplet-bsed energy genaration from 3D printed substrate, a self-powered 3D printed house is designed. It shows energy harvesting ability from rainwater as schematically ptresented in **Figure 4E**. The ultrathin Gd$_2$Te$_3$ was coated on the floor's roof (**Figure S8**). The output voltage is observed under continuous dropping of artifical rainwater, (**Figure 4F, Figure S8** and **Movie S2)**. Here, the self- powered 3DP house is an immediate application of energy harvesting from natural resources. Therefore, it could be further integrated with other objects for energy harvesting from rainwater.

To get insight into the voltage generation on the Gd$_2$Te$_3$ surface, we performed first-principles calculations to obtain the electron density difference and conductivity. Bulk Gd$_2$Te$_3$ structure was taken from the Materials Project Database [33]. The optimized lattice parameters for Gd$_2$Te$_3$ were found to be *a*=12.47 Å, *b*=4.83 Å, *c*=12.40 Å. Next, three layers (112) surface of $Gd_2Te_3$ were chosen for further calculation since it has the most prominent peak in the experimental XRD plot (**Figure 5A**). Lattice parameter of (112) surface was $a$=13.25 Å, $b$=15.30 Å, $c$=26.80 Å. To understand the charge distribution, we did a Bader charge analysis of the (112) surface, and it shows that Te atoms have a negative Bader charge, thereby accepting electrons from Gd atoms having a positive Bader charge. Hence, the considered surface accumulates both negative and positive polarities. However, their concentration of the charges is not the same on the surface. In order to understand the predominating polarity present on the surface, we have calculated the charge accumulation to visualize the charge redistribution on a surface:

$\Delta\rho = \rho(Gd_2Te_3) - \rho(Gd_2) - \rho(Te_3)$,  … (1)

where $\rho(Gd_2Te_3)$, $\rho(Gd_2)$, $\rho(Te_3)$ are the charge densities of an isolated system of $Gd_2Te_3$, $Gd_2$, $Te_3$, respectively. The charge accumulation-depletion plot (**Figure 5B**), shows that the (112) surface of $Gd_2Te_3$ has more positively polarized regions than the negatively polarized regions. This is the reason we got a positive zeta potential (+4.8 mV) in the experimental measurement. Furthermore, to confirm this, electrical conductivity calculations were performed for the chosen surface to quantify the potential of $Gd_2Te_3$ to generate electricity.

Utilizing the electronic structure of the system, the electronic transport properties have been calculated by solving the semi-classical Boltzmn transport equation within constant scattering time approximation (CSTA), as implemented in Boltztrap code [34]. A dense Monkhorst-Pack k-point (21 × 21 × 1) has been used for electronic transport calculations. Calculated thermopower and electrical conductivity scaled by relaxation time are shown in **Figure 5C-D**. At room temperature, within a carrier concentration range of $10^{12}$- $10^{13}$ cm$^{-2}$, a wide range of Seebeck coefficient values from 300-400 µV/K have been observed for both p- and n-type charge carriers. This suggests that a large amount of voltage can be generated with a given temperature gradient in this system. It also confirms the contribution of the charge carrier in the droplet and surface. Additionally, in the similar carrier concentration range, at room temperature, the scaled electrical conductivity ($\sigma/\tau$) exhibits a maximum of 2-4×10$^{17}$ $\Omega^{-1}$m$^{-1}$s$^{-1}$ for p-type charge carriers. In contrast, n-type charge carriers possess a lower $\sigma/\tau$. Considering the constant relaxation time approximation (CRTA), i.e, with $a\tau$ = 10 fs, the maximum conductivity for p-type charge carriers is 2000 S/m, which is reasonably high conductivity, implying the potential of 2D Gd$_2$Te$_3$ to generate electricity [35]. Therefore, as the electrical conductivity increases with the droplet, the attraction and release of charge play an important role in generating electricity. Based on the theoretical calculation, the detailed mechanism of charge transfers between NaCl solution and Gd$_2$Te$_3$ is schematically shown in **Figure 5E**. As seen from the experimental results, the ions strongly affect the generation of charge in the sample. The concentration of charges in the solution plays a key role in electrification, as we have seen in the previous result. The free ions in DI water are meager; therefore, DI droplets cannot generate much output voltage by contact electrification. However, the output voltage increases due to the electrification process after increasing free ions in the DI water by adding NaCl salt, as shown in **Figure 5E**. Due to the positive surface charge of the Gd$_2$Te$_3$, negatively Cl$^-$ ions are attracted. As the droplet moves on the substrate, the contact area increases. The contact between the ionic droplet and Gd$_2$Te$_3$ causes more charge transfer. During the separation of the droplets, as the area of droplets connected to the Gd$_2$Te$_3$ gradually decreases, the concentration of positive charges increases due to the mutual attraction of the positive and negative charges. The increased charge density results in a stronger electric field. From electro-kinetic theory, it is known that when ionic solutions are brought in contact with solid surfaces, some of the ions adsorb onto the surface due to electrochemical interactions, forming a layer of either cations or anions depending on the surface charge of the material. To balance this out, the charge accumulates on the solid surface [7][36].

Here, ultrathin Gd$_2$Te$_3$ has huge positive surface charges, as found from theory and experimental results. Additionally, the layered nature of the material-coated 3D printed porous structure provides sufficient surface area for the ions from the solution to adsorb. Therefore, positively charged 2D material is adsorbed negative Cl$^-$ ions on the surface of the 3D printed cell. To balance this negatively charged layer of Cl$^-$ ions, positive charges accumulate near the solid-liquid interface as the electrons move to the opposite side of the 2D material, towards the Gd$_2$Te$_3$-polymer interface, and it suggests that charge transfer is crucial for the generation of overall output voltage. As mentioned earlier, the manipulation of output performance of the device using a graphite rod can be understood as follows. In the presence of a rubbing graphite rod, the graphene layer represents the steady and dynamic states of the NaCl droplet on the Gd$_2$Te$_3$ surface. Considering the different adsorption capacities of ions (Na$^+$, Cl$^-$) on Gd$_2$Te$_3$ and graphene layer, Cl$^-$ ions are adsorbing on Gd$_2$Te$_3$, while Na$^+$ ions are adsorbing on the graphene surface [7] [36]. The graphene surface and vice versa repel the negative ions; therefore, two different electron transfer interfaces were formed, as shown by square boxes in the figure (**Figure 3I**) [37]. Thus, the surface charge of the droplet can be manipulated by controlling the movement of the negatively charged graphene surface. As the graphite rod touches the droplet surface, the system (2D Gd$_2$Te$_3$+droplet+graphen sheets) gives output voltage like a soft micro-robot cover upon the substrate. Therefore, contact electrification with a surface charge of the droplet can be used directly in several applications combined with lightweight flexible 3D printed substrates.

## 3. Conclusions

In summary, we have reported electricity generation using ultrathin 2D Gd$_2$Te$_3$ decorated 3D printed substrate as an active material. The electricity is generated directly during the movement of ionic liquid on 2D Gd$_2$Te$_3$ due to electrostatic charge transfer at the droplet boundary via contact electrification. We have also examined the output performance as a function of height, and it shows a linear increment with height. More importantly, the energy generation capability of Gd$_2$Te$_3$ decorated porous objects is further successfully enhanced with random porosity structures. The DFT calculations confirm the presence of predominant positive polarity over the surface of $Gd_2Te_3$ that interacts with a solution to generate electricity with a wide range of Seebeck coefficients. This is further supported by the calculated high electrical conductivity of $2\text{-}4\times10^{17}$ $\Omega^{-1}m^{-1}s^{-1}$ for p-type charge carriers. Additionally, the output voltage was manipulated using a graphite rod by charge accumulations. Therefore, coupling ultrathin 2D materials and 3D printing technology can be a good choice for various applications in energy-harvesting, biomedical sector, sensors, hydrodynamic sensors, etc.

## 4. Experimental Section

### 4.1 Synthesis:

Bulk $Gd_2Te_3$ was fabricated using the vacuum induction melting method. Metal samples of 99.99% purity of Gd and Te were used for preparing the compound containing 4.7 g of Gd and 5.3 g of Te. The alloys were prepared by melting the constituent element at a temperature of 1250°C in a quartz tube using vacuum induction melting in an argon atmosphere and allowed to cool naturally in Ar. 2D sheets of $Gd_2Te_3$ were synthesized using ultrasonication-assisted liquid phase exfoliation with 2-propanal (IPA) as solvent. 1.5 g of powdered $Gd_2Te_3$ was ultrasonicated at 30 kHz in 300 mL of IPA for 30 minutes. Using 2-Propanol (isopropyl alcohol or IPA) for exfoliation was primarily driven by its low toxicity compared to other solvents and its high volatility, which allows to deposit 2D sheets on surfaces.

### 4.2 Preparation of porous 3D printed structure

The 15 x 30 x 0.08 mm specification of structure were fabricated using polylactic acid filament of 1.75 mm (with tolerance ± 0.1) diameter via extrusion-based Flashforge adventure 3 printer. The porous structures printed at extruder temperature of 210°C and platform temperature of 50°C with print speed of 40 mm/s and travel speed of 70 mm/s. The infill density kept at 60% for hexagonal, triangular and line porosity whereas for random porosity it was kept at 100%.

All the structures are printed at hyper resolution (single layer thickness of 80 μm) and the solidification of post-printed structures were carried out with the help of forced fan cooling.

**4.3 Characterizations:**

X-ray diffraction (XRD) patterns of the crystalline phases of the drop-casted 2D $Gd_2Te_3$ were obtained by X-ray diffractometer (Bruker, D8 Advance) with a Cu-Kα (λ = 1.5406Å) radiation source, operating at 40 kV voltage and 40 mA current. Scanning electron microscopy (SEM-Jeol JSM-IT300HR) operated at an acceleration voltage of 20 kV and 7.475 nA was used to image both bulk and 2D $Gd_2Te_3$. Transmission electron microscopy (TEM-JEM 2100 HRTEM) and Atomic force microscope (AFM, Nanosurf easy scan 2) was used to image the 2D sheets. Attached energy dispersion spectrometer units to both SEM and TEM were used to get a composition of both bulk and 2D $Gd_2Te_3$. The particle size distribution and zeta potential were obtained from Horiba Scientifica Nano Particle Analyzer SZ-100. Raman spectroscopy was done using WITec UHTS Rama Spectrometer (WITec, UHTS 300 VIS, Germany) at the laser excitation wavelength of 532 nm at Room Temperature (RT). The optical absorbance property of pristine 2D $Gd_2Te_3$ was characterized using a UV-Vis spectrometer from 200 to 800 nm. The output voltage of the hybrid nanogenerator was measured by a digital storage oscilloscope (DSO, Tektronix, TBS1072B).

**4.4 Computational Methodology:**

Density functional theory (DFT) was done with the Vienna ab initio simulations (VASP) package.[38] The Electron-ion interactions were described using the all-electron projector augmented wave pseudopotentials [39], and Perdew-Bruke-Ernzehof (PBE) generalized gradient approximation (GGA) [40] was used to approximate the electronic exchange correlations with an effective on-site Hubbard ($U_{eff}$ = U-J = 6 eV) parameter was used for the Gd- states in DFT+U method as introduced by Dudarev et al.[41] The plane-wave kinetic energy cut-off of 520 eV was used. All the structures were optimized using a conjugate gradient scheme until the energies and the components of forces reached $10^{-6}$ eV and 0.001 eV Å$^{-1}$ for bulk Gd$_2$Te$_3$ and $10^{-5}$ and 0.01 eV Å-1 for (112) surface of Gd$_2$Te$_3$, respectively. A vacuum of 10 Å was added in the z-direction to prevent interactions between the periodic images. The Brillouin zone of bulk Gd$_2$Te$_3$ and (112) surface of Gd$_2$Te$_3$ were sampled with 7×7×7 Monkhorst-Pack and 3×3×1 Monkhorst-Pack, respectively. All the calculations were spin-polarized.

**Acknowledgment**

**Declaration of Competing Interest:** There are no conflicts to declare.

**REFERENCES**

[1] S. Chu, A. Majumdar, Opportunities and challenges for a sustainable energy future, Nature. 488 (2012) 294–303. https://doi.org/10.1038/nature11475.

[2] J. Xiong, M.F. Lin, J. Wang, S.L. Gaw, K. Parida, P.S. Lee, Wearable All-Fabric-Based Triboelectric Generator for Water Energy Harvesting, Adv. Energy Mater. 7 (2017) 1–10. https://doi.org/10.1002/aenm.201701243.

[3] J. Nie, Z. Ren, L. Xu, S. Lin, F. Zhan, X. Chen, Z.L. Wang, Probing Contact-Electrification-Induced Electron and Ion Transfers at a Liquid–Solid Interface, Adv. Mater. 32 (2020) 1–11. https://doi.org/10.1002/adma.201905696.

[4] S. Nie, H. Guo, Y. Lu, J. Zhuo, J. Mo, Z.L. Wang, Superhydrophobic Cellulose Paper-Based Triboelectric Nanogenerator for Water Drop Energy Harvesting, Adv. Mater. Technol. (2020) 2000454. https://doi.org/10.1002/admt.202000454.

[5] L. Zhao, L. Liu, X. Yang, H. Hong, Q. Yang, J. Wang, Q. Tang, Cumulative charging behavior of water droplet driven freestanding triboelectric nanogenerators toward hydrodynamic energy harvesting, J. Mater. Chem. A. 8 (2020) 7880–7888. https://doi.org/10.1039/d0ta01698e.

[6] Z.L. Wang, T. Jiang, L. Xu, Toward the blue energy dream by triboelectric nanogenerator networks, Nano Energy. 39 (2017) 9–23. https://doi.org/10.1016/j.nanoen.2017.06.035.

[7] J. Yin, X. Li, J. Yu, Z. Zhang, J. Zhou, W. Guo, Generating electricity by moving a droplet of ionic liquid along graphene, Nat. Nanotechnol. 9 (2014) 378–383. https://doi.org/10.1038/nnano.2014.56.

[8] N. Li, C. Yu, Z. Dong, L. Jiang, Finger directed surface charges for local droplet

motion, Soft Matter. 16 (2020) 9176–9182. https://doi.org/10.1039/d0sm01073a.

[9] J.A. Lv, Y. Liu, J. Wei, E. Chen, L. Qin, Y. Yu, Photocontrol of fluid slugs in liquid crystal polymer microactuators, Nature. 537 (2016) 179–184. https://doi.org/10.1038/nature19344.

[10] C. Zhu, T. Liu, F. Qian, T.Y.J. Han, E.B. Duoss, J.D. Kuntz, C.M. Spadaccini, M.A. Worsley, Y. Li, Supercapacitors Based on Three-Dimensional Hierarchical Graphene Aerogels with Periodic Macropores, Nano Lett. 16 (2016) 3448–3456. https://doi.org/10.1021/acs.nanolett.5b04965.

[11] K. Ghosh, M. Pumera, Free-standing electrochemically coated MoS: Xbased 3D-printed nanocarbon electrode for solid-state supercapacitor application, Nanoscale. 13 (2021) 5744–5756. https://doi.org/10.1039/d0nr06479c.

[12] R.S. Ambekar, B. Kushwaha, P. Sharma, F. Bosia, M. Fraldi, N.M. Pugno, C.S. Tiwary, Topologically engineered 3D printed architectures with superior mechanical strength, Mater. Today. 48 (2021) 72–94. https://doi.org/10.1016/j.mattod.2021.03.014.

[13] E. García-T̂On, S. Barg, J. Franco, R. Bell, S. Eslava, E. D'Elia, R.C. Maher, F. Guitian, E. Saiz, Printing in three dimensions with Graphene, Adv. Mater. 27 (2015) 1688–1693. https://doi.org/10.1002/adma.201405046.

[14] R.S. Ambekar, E.F. Oliveira, B. Kushwaha, V. Pal, P.M. Ajayan, A.K. Roy, D.S. Galvao, C.S. Tiwary, Flexure resistant 3D printed zeolite-inspired structures, Addit. Manuf. 47 (2021) 102297. https://doi.org/10.1016/j.addma.2021.102297.

[15] B. Kushwaha, K. Dwivedi, R.S. Ambekar, V. Pal, D.P. Jena, D.R. Mahapatra, C.S. Tiwary, Mechanical and Acoustic Behavior of 3D-Printed Hierarchical Mathematical Fractal Menger Sponge, Adv. Eng. Mater. 23 (2021) 2001471. https://doi.org/10.1002/adem.202001471.

[16] R.S. Ambekar, I. Mohanty, S. Kishore, R. Das, V. Pal, B. Kushwaha, A.K. Roy, S. Kumar Kar, C.S. Tiwary, Atomic Scale Structure Inspired 3D-Printed Porous Structures with Tunable Mechanical Response, Adv. Eng. Mater. 23 (2021) 2001428. https://doi.org/10.1002/adem.202001428.

[17] L.M. Schoop, M.N. Ali, C. Straßer, A. Topp, A. Varykhalov, D. Marchenko, V. Duppel, S.S.P. Parkin, B. V. Lotsch, C.R. Ast, Dirac cone protected by non-symmorphic symmetry and three-dimensional Dirac line node in ZrSiS, Nat. Commun. 7 (2016). https://doi.org/10.1038/ncomms11696.

[18] D.A. Zocco, J.J. Hamlin, K. Grube, J.H. Chu, H.H. Kuo, I.R. Fisher, M.B. Maple, Pressure dependence of the charge-density-wave and superconducting states in GdTe3, TbTe3, and DyTe3 pressure dependence of the charge-density- ⋯ Zocco, Hamlin, Grube, Chu, Kuo, Fisher, and Maple, Phys. Rev. B - Condens. Matter Mater. Phys. 91 (2015) 1–7. https://doi.org/10.1103/PhysRevB.91.205114.

[19] S. Siddique, C. Chowde Gowda, S. Demiss, R. Tromer, S. Paul, K.K. Sadasivuni, E.F. Olu, A. Chandra, V. Kochat, D.S. Galvão, P. Kumbhakar, R. Mishra, P.M. Ajayan, C. Sekhar Tiwary, Emerging two-dimensional tellurides, Mater. Today. (2021). https://doi.org/10.1016/j.mattod.2021.08.008.

[20] S. Siddique, C.C. Gowda, R. Tromer, S. Demiss, A.R.S. Gautam, O.E. Femi, P. Kumbhakar, D.S. Galvao, A. Chandra, C.S. Tiwary, Scalable Synthesis of Atomically Thin Gallium Telluride Nanosheets for Supercapacitor Applications, ACS Appl. Nano Mater. 4 (2021) 4829–4838. https://doi.org/10.1021/acsanm.1c00428.

[21] A.S. Aji, R. Nishi, H. Ago, Y. Ohno, High output voltage generation of over 5 V from liquid motion on single-layer MoS2, Nano Energy. 68 (2020) 104370. https://doi.org/10.1016/j.nanoen.2019.104370.

[22] J.S. Swinnea, H. Steinfink, L.R. Danielson, Crystal chemistry and thermoelectric properties of Gd2Te3, J. Appl. Crystallogr. 20 (1987) 102–104. https://doi.org/10.1107/S0021889887087065.

[23] I.P. Muthuselvam, R. Nehru, K.R. Babu, K. Saranya, S.N. Kaul, S.M. Chen, W.T. Chen, Y. Liu, G.Y. Guo, F. Xiu, R. Sankar, Gd2Te3: An antiferromagnetic semimetal, J. Phys. Condens. Matter. 31 (2019). https://doi.org/10.1088/1361-648X/ab1570.

[24] P. Arun, A.G. Vedeshwar, On the structure of stibnite (Sb2S3), J. Mater. Sci. 31 (1996) 6507–6510. https://doi.org/10.1007/BF00356255.

[25] C.F. Holder, R.E. Schaak, Tutorial on Powder X-ray Diffraction for Characterizing Nanoscale Materials, ACS Nano. 13 (2019) 7359–7365. https://doi.org/10.1021/acsnano.9b05157.

[26] S. Lei, J. Lin, Y. Jia, M. Gray, A. Topp, G. Farahi, S. Klemenz, T. Gao, F. Rodolakis, J.L. McChesney, C.R. Ast, A. Yazdani, K.S. Burch, S. Wu, N.P. Ong, L.M. Schoop, High mobility in a van der Waals layered antiferromagnetic metal, Sci. Adv. 6 (2020) 1–10. https://doi.org/10.1126/sciadv.aay6407.

[27] Q. Lai, S. Zhu, X. Luo, M. Zou, S. Huang, Ultraviolet-visible spectroscopy of graphene oxides, AIP Adv. 2 (2012) 2–7. https://doi.org/10.1063/1.4747817.

[28] Y. Zhang, Q. Tang, B. He, P. Yang, Graphene enabled all-weather solar cells for electricity harvest from sun and rain, J. Mater. Chem. A. 4 (2016) 13235–13241. https://doi.org/10.1039/C6TA05276B.

[29] D. Park, S. Won, K.-S. Kim, J.-Y. Jung, J.-Y. Choi, J. Nah, The influence of substrate-dependent triboelectric charging of graphene on the electric potential generation by the flow of electrolyte droplets, Nano Energy. 54 (2018) 66–72. https://doi.org/10.1016/j.nanoen.2018.09.054.

[30] H. Li, D. Zhang, H. Wang, Z. Chen, N. Ou, P. Wang, D. Wang, X. Wang, J. Yang, Molecule-Driven Nanoenergy Generator, Small. (2018) 1804146. https://doi.org/10.1002/smll.201804146.

[31] S. Ho Lee, D. Kim, S. Kim, C.-S. Han, Flow-induced voltage generation in high-purity metallic and semiconducting carbon nanotubes, Appl. Phys. Lett. 99 (2011) 104103. https://doi.org/10.1063/1.3634209.

[32] H. Zhong, J. Xia, F. Wang, H. Chen, H. Wu, S. Lin, Graphene-Piezoelectric Material Heterostructure for Harvesting Energy from Water Flow, Adv. Funct. Mater. 27 (2017) 1604226. https://doi.org/10.1002/adfm.201604226.

[33] A. Jain, S.P. Ong, G. Hautier, W. Chen, W.D. Richards, S. Dacek, S. Cholia, D. Gunter, D. Skinner, G. Ceder, K.A. Persson, Commentary: The Materials Project: A materials genome approach to accelerating materials innovation, APL Mater. 1 (2013) 011002. https://doi.org/10.1063/1.4812323.

[34] G.K.H. Madsen, D.J. Singh, BoltzTraP. A code for calculating band-structure dependent quantities, Comput. Phys. Commun. 175 (2006) 67–71. https://doi.org/10.1016/j.cpc.2006.03.007.

[35] D. Sheberla, L. Sun, M.A. Blood-Forsythe, S. Er, C.R. Wade, C.K. Brozek, A. Aspuru-Guzik, M. Dincă, High Electrical Conductivity in Ni 3 (2,3,6,7,10,11-hexaiminotriphenylene) 2 , a Semiconducting Metal–Organic Graphene Analogue, J. Am. Chem. Soc. 136 (2014) 8859–8862. https://doi.org/10.1021/ja502765n.

[36] S.S. Kwak, S. Lin, J.H. Lee, H. Ryu, T.Y. Kim, H. Zhong, H. Chen, S.W. Kim, Triboelectrification-Induced Large Electric Power Generation from a Single Moving Droplet on Graphene/Polytetrafluoroethylene, ACS Nano. 10 (2016) 7297–7302. https://doi.org/10.1021/acsnano.6b03032.

[37] J. Xiong, M.-F. Lin, J. Wang, S.L. Gaw, K. Parida, P.S. Lee, Wearable All-Fabric-Based Triboelectric Generator for Water Energy Harvesting, Adv. Energy Mater. 7 (2017) 1701243. https://doi.org/10.1002/aenm.201701243.

[38] G. Kresse, J. Hafner, Ab initio molecular dynamics for liquid metals, Phys. Rev. B. 47 (1993) 558–561. https://doi.org/10.1103/PhysRevB.47.558.

[39] G. Kresse, D. Joubert, From ultrasoft pseudopotentials to the projector augmented-wave method, Phys. Rev. B. 59 (1999) 1758–1775. https://doi.org/10.1103/PhysRevB.59.1758.

[40] J.P. Perdew, K. Burke, M. Ernzerhof, Generalized Gradient Approximation Made Simple, Phys. Rev. Lett. 77 (1996) 3865–3868. https://doi.org/10.1103/PhysRevLett.77.3865.

[41] S.L. Dudarev, G.A. Botton, S.Y. Savrasov, C.J. Humphreys, A.P. Sutton, Electron-energy-loss spectra and the structural stability of nickel oxide: An LSDA+U study, Phys. Rev. B. 57 (1998) 1505–1509. https://doi.org/10.1103/PhysRevB.57.1505.

**Figures**

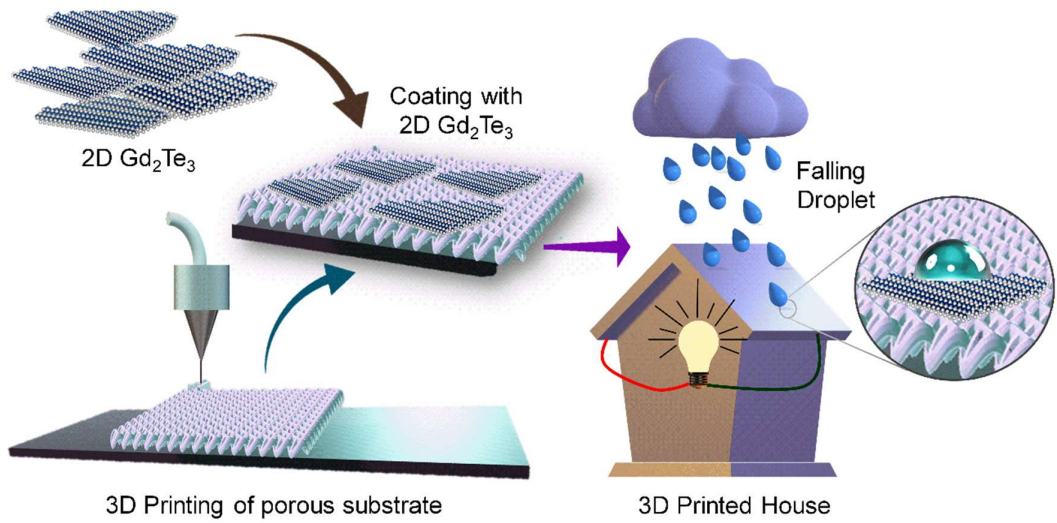

**Figure 1:** *Schematic representation of fabrication of 3D printed devices decorated with ultrathin $Gd_2Te_3$ and energy generation using liquid droplets.*

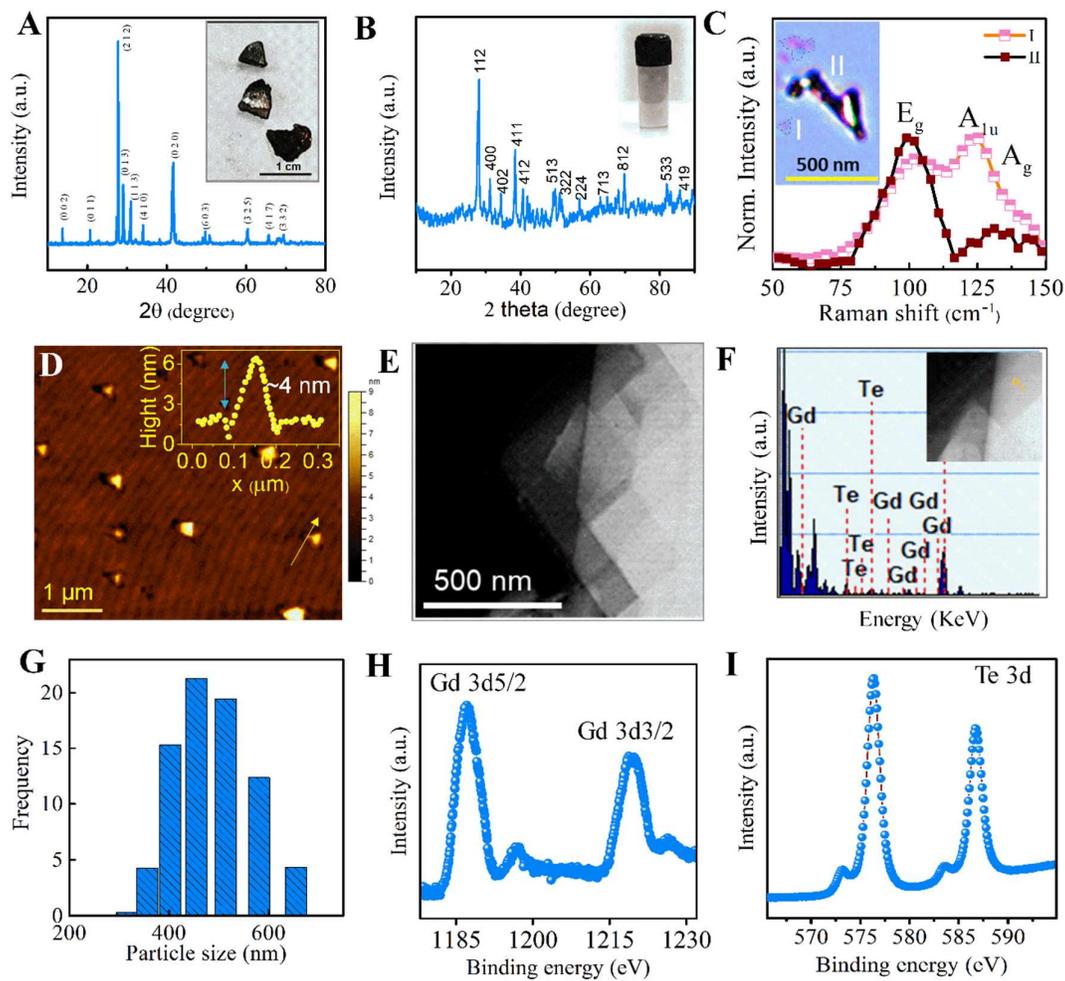

***Figure 2.*** *(A) XRD pattern of bulk $Gd_2Te_3$. Inset shows the digital photo of as-synthesized bulk $Gd_2Te_3$. (B). XRD peaks of 2D $Gd_2Te_3$. Inset shows the dispersion of 2D $Gd_2Te_3$ in IPA after liquid exfoliation. (C) Raman spectra of 2D $Gd_2Te_3$ of varying thickness, (D) AFM and line profile of 2D $Gd_2Te_3$ drop-casted on a Si substrate. (E) Bright-field TEM image of the exfoliated sheets, (F) An EDX spectrum is showing Gd and Te. (G) Average particle size analysis of exfoliated samples. The XPS spectrum of the 2D $Gd_2Te_3$ with fitted peaks for (H) Gd 3d and (I) Te 3d.*

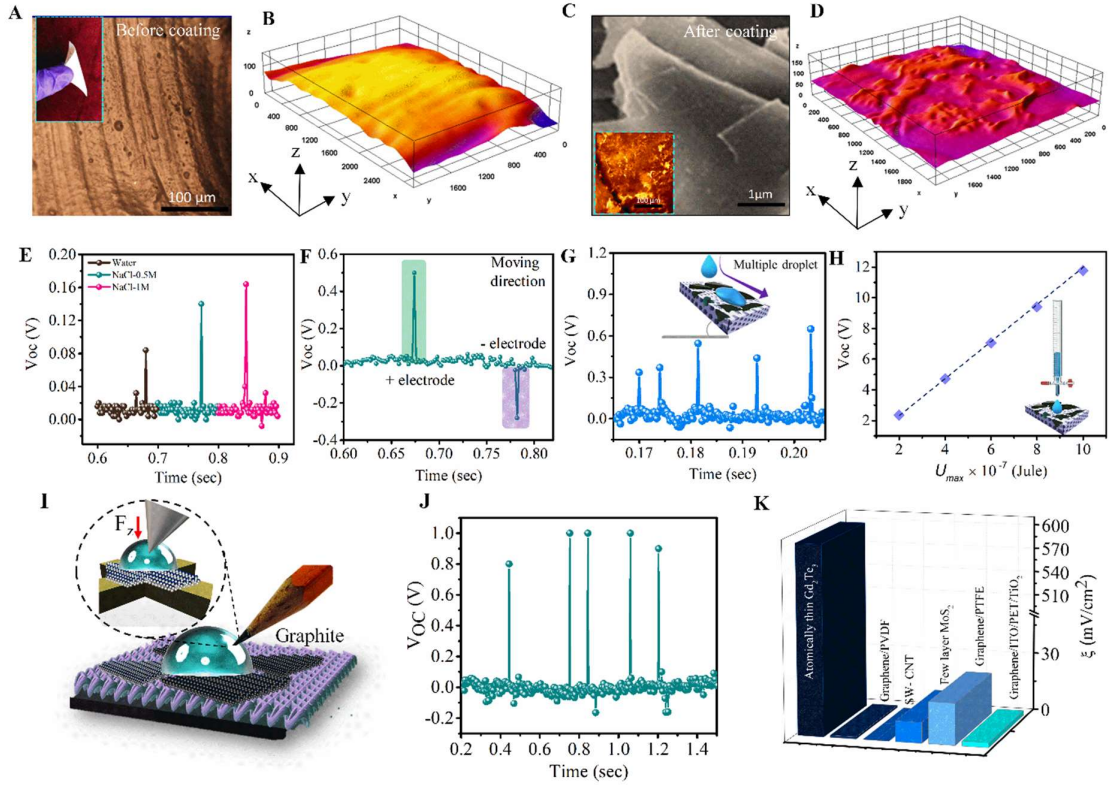

***Figure 3:*** *Optical microscopy image and SEM image of 3D printed substrate (A) before and (C) after decoration with 2D material. The inset of (A) shows the digital photograph of the flexible substrate. 3D surface analysis of the printed structure (B) before and (D) after decoration. (E) Droplets of different ionic concentrations dropped from the same height. (F) The pulse is generated along the moving direction of the droplet. (G) Multiple droplets are generating output voltage. Inset shows the schematic of the measurements setup. (H) Output voltage increase vs. potential energy at the surface by falling droplets from varying heights. (I) Schematic representation of NaCl droplet on cell and moving direction, graphite charge layer. (J) The output voltage of 3D printed cell. (K) Comparisons of the liquid droplet-based energy generation with other 2D materials.*

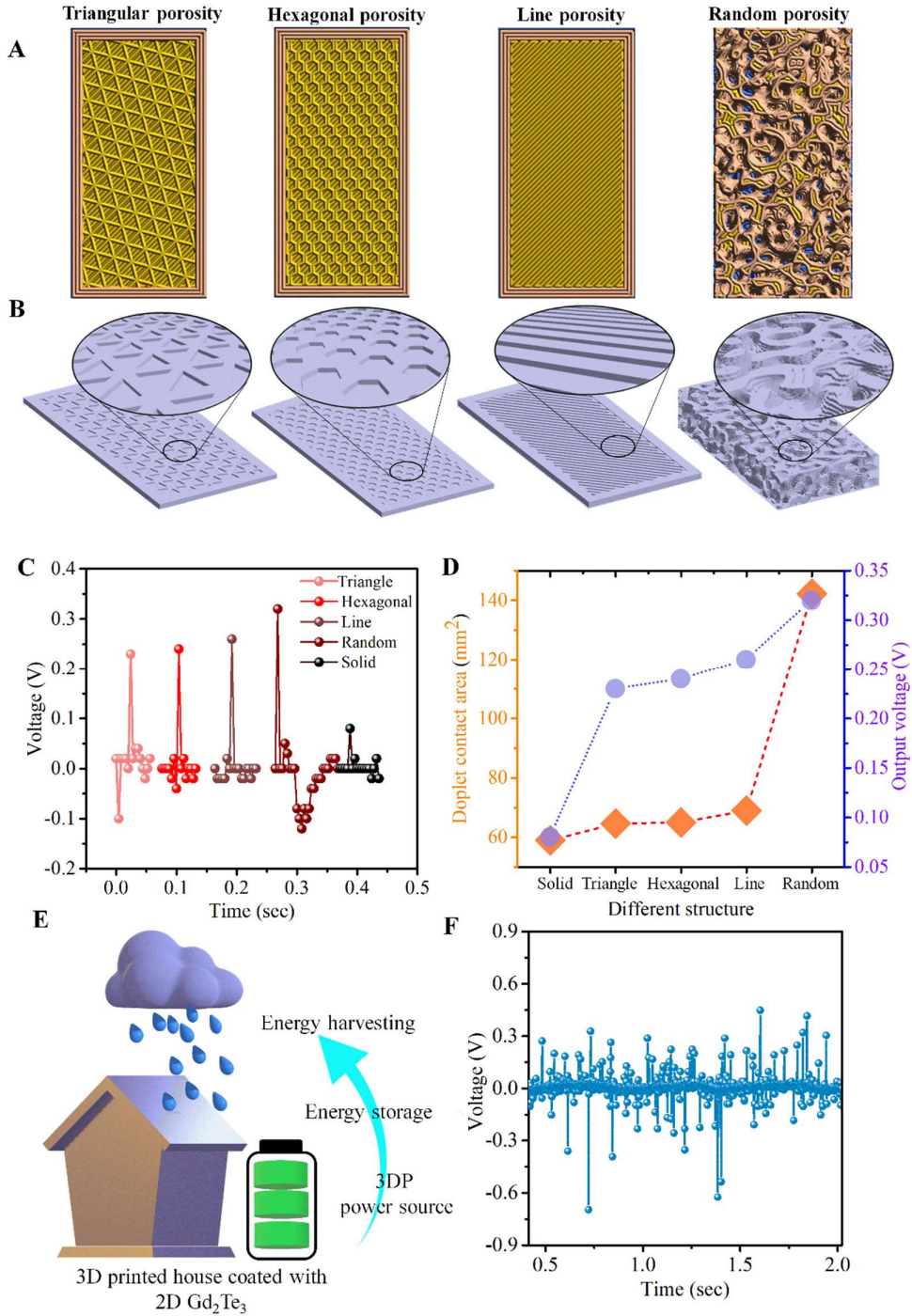

***Figure 4:*** *(A) Top view and (B) Isometric view of 3D printed porous structures (insect show zoomed region of pores). (C) Effect of copious porosity on the output voltage. (D) Effect of water droplet contact area on output voltage. (E) 3D printed smart house for energy harvesting from raindrops. (F) Output voltage generated from artificial raindrops from 3DP-house.*

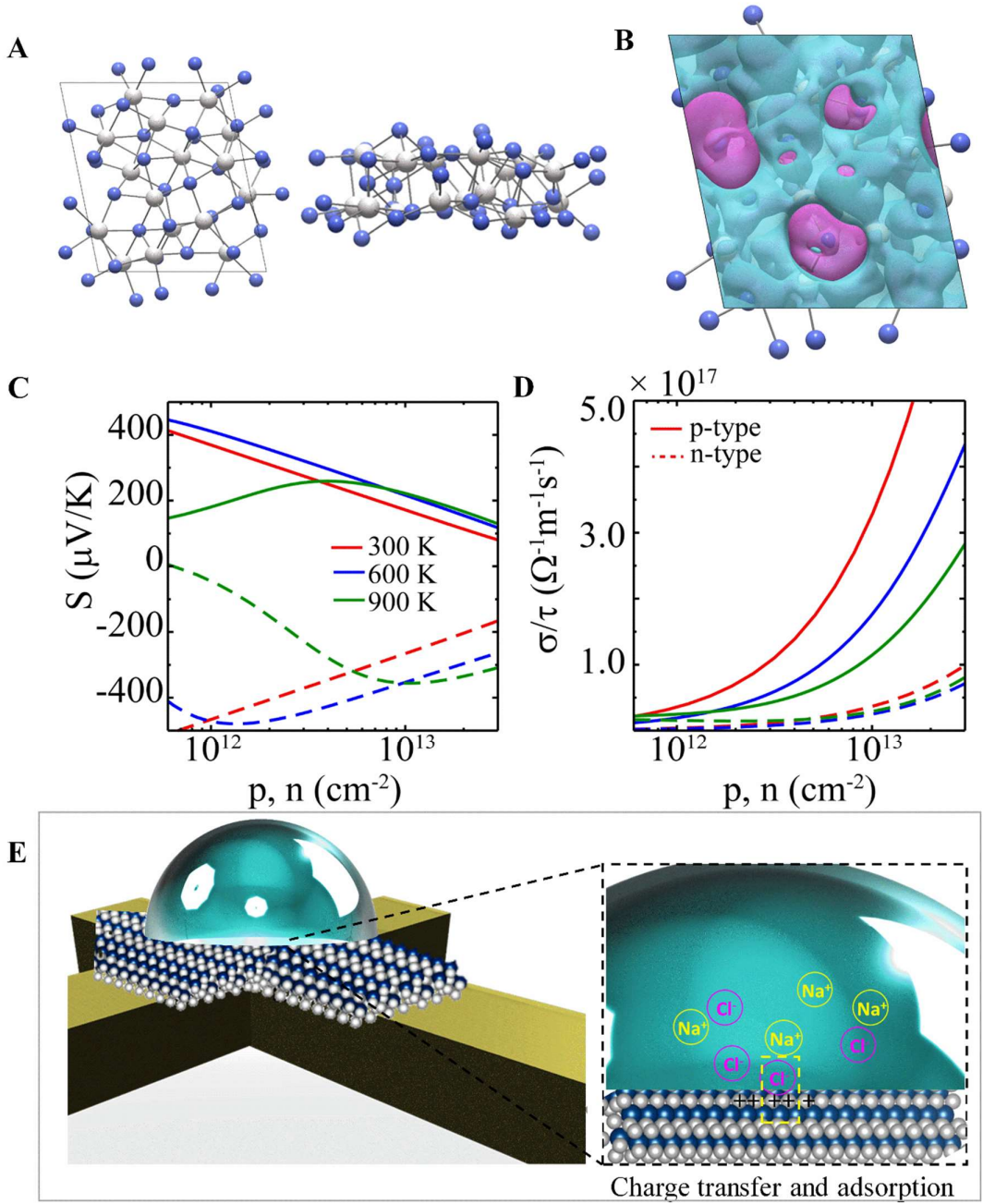

*Figure 5*: *(A) (112) surface of Gd$_2$Te$_3$, on-top view, and side-on view. Grey balls represent Gd atoms and blue balls represent Te atoms. (B) Charge redistribution diagram. Sky blue represents electron depleted region and pink represents the electron accumulated region. (C) Seebeck coefficient vs carrier concentration diagram, (D) Electronic conductivity vas carrier concentration diagram. (E) Schematic diagram of charge generation mechanism due to contact electrification.*

**Graphical Abstract**

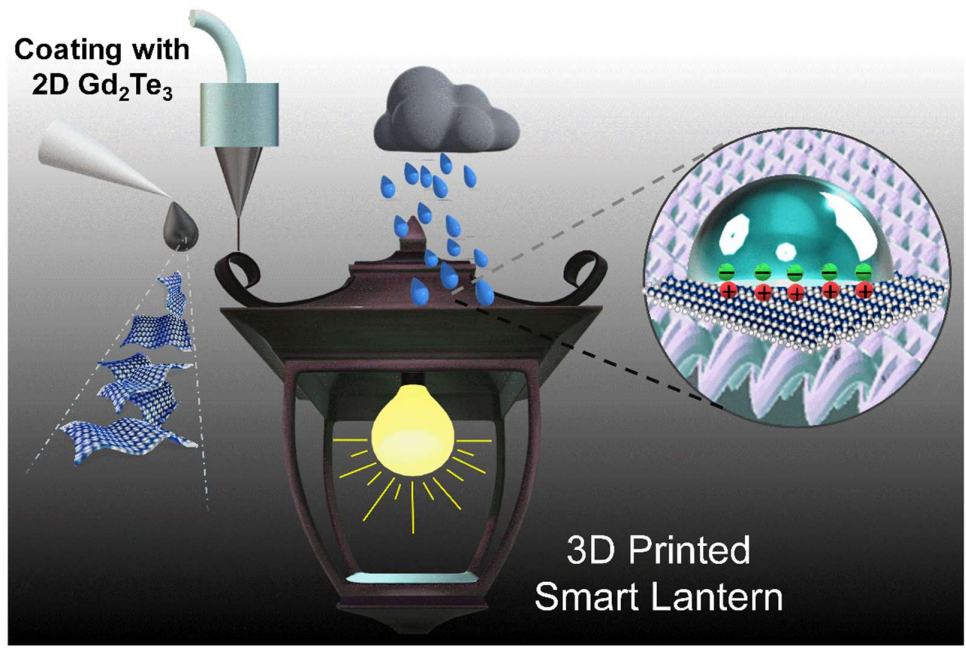